\documentclass[twocolumn,showpacs,prl]{revtex4}
\usepackage{amsmath}
\usepackage{amssymb}
\usepackage{graphicx}
\usepackage{dcolumn}
\usepackage{bm}

\begin{document}

\title{Intra-Landau-level collective excitations in a bilayer disordered
electronic system}
\author{Samvel M. Badalyan}
\altaffiliation[Permanent address: ]{Radiophysics Department, Yerevan State University, Yerevan, 375025 Armenia}
\email{badalyan@lx2.yerphi.am}
\affiliation{Department of Physics, Chonnam National University, Kwangju 500-757, Korea}
\author{Chang Sub Kim}
\email{cskim@boltzmann.chonnam.ac.kr}
\affiliation{Department of Physics, Chonnam National University, Kwangju 500-757, Korea}
\date{\today }

\begin{abstract}
We investigate intra-Landau-level collective excitations in a bilayer
disordered two-dimensional electron system exposed to a perpendicular
magnetic field. The energy spectrum is calculated within the random phase
approximation by taking into account electron-impurity scattering in the
self-consistent Born approximation which includes consistent vertex
corrections. Signatures of these bilayer excitations in drag and collective
excitation measurements are identified.
\end{abstract}

\maketitle

%\preprint{APS/123-QED}

%Lines break %automatically or can be forced with \\







%\preprint{APS/123-QED}

%Lines break %automatically or can be forced with \\




In bilayer two dimensional electron systems (2DES), an additional degree of
freedom, associated with the layer index, produces collective excitations
(CE), which have no counterpart in the individual 2DES \cite{eisensuen}. New
interlayer electron correlations arise which govern different ferromagnetic
and antiferromagnetic phases of the quantum Hall states \cite{isospin}. The
bilayer CE also play an essential role in the dynamical screening of
interlayer Coulomb interaction in the frictional drag effect both in zero
\cite{flensberg,hill97,noh98} and finite magnetic fields \cite%
{hill97,wu,khaetskii,manolescu,ssc}.

The basic source of physical information on the fundamental properties of a
system is the spectrum of CE. In zero magnetic field, the spectrum of CE in
the bilayer 2DES without disorder effects has been studied comprehensively
both in theory \cite{sarmachaplik}\ and experiment \cite{kainth}. The
spectrum includes not only the optical wave with in-phase density
fluctuations and the usual square-root dispersion, but also an additional
acoustical branch with out-of-phase fluctuations and linear dispersion. In
the finite magnetic fields, the dispersion of inter-Landau level (LL) CE
with energies close to the multiples of the cyclotron energy was obtained
analytically at the zero temperature limit \cite{aizin}. Experimental
measurements of the spectrum of inter-LL\ CE in bilayer 2DES were performed
only recently: the principal Bernstein magnetoplasmon modes were observed by
means of inelastic light scattering \cite{tovstonog}.

Of central importance, however, is the spectrum of CE in the extreme quantum
limit when no kinetic energy change occurs in the 2DES. These CE with
energies, $\hbar \omega $, below the cyclotron energy, $\hbar \omega _{B},$
are referred to as the intra-LL CE. In the fractional quantum Hall effect
(FQHE) regime the intra-LL CE are the well-known magnetorotons in single
\cite{girvin} and double layer \cite{rennamd94} 2DES with the dispersion
minima at wave vectors $q\sim 1/\ell _{B}$ \cite{kallin}, $\ell _{B}$ is the
magnetic length. In disordered 2DES, a continuum of particle-hole
excitations takes place within a broadened LL. In the individual disordered
2DES the spectrum of intra-LL CE has been studied by Antoniou \textit{et al}%
. \cite{anton} for the lowest LL and it has been found that the plasmon-like
CE emerge above the particle-hole continuum (PHC). It has been also
predicted by Merkt \cite{merkt} that in the disordered 2DES,
electron-electron (e-e) interaction overcomes Kohn's theorem \cite{kohn} and
results in a low-lying disorder-induced cyclotron resonance mode. Recently,
the importance of intra-LL CE in the drag effect has been also discussed
\cite{manolescu}.

In the present paper we calculate the spectrum of intra-LL CE in the bilayer
disordered 2DES in a perpendicular magnetic field at finite temperatures
when a higher LL is partially filled. We treat e-e interaction within the
random phase approximation (RPA) and electron-impurity (e-i) scattering in
the self-consistent Born approximation (SCBA) \cite{ando}, which includes
the consistent vertex corrections to the response function of the individual
2DES \cite{anton}. We obtain that the whole spectrum of intra-LL CE is
located in a finite wave vector interval and, depending on the system
parameters, there can exist $l_{0}$ forbidden zones within this wave vector
interval ($l_{0}$ is the Landau index of the outermost partially filled LL).
Although our approach allows us to take into account coupling of the
broadened LL at finite temperatures, the coupling is weak in the regime of
our interest here. Therefore, the response function becomes small and the
2DES unable to respond in the forbidden wave vector zones around $q_{r}$
where $q_{r}$ are determined from the zeros of the Laguerre polynomial, $%
L_{l_{0}}(t_{r})=0$ and $t_{r}=(q_{r}\ell _{B})^{2}/2$ ($r=1,2,...,l_{0}$).
In each allowed wave vector zone, we obtain the spectrum of bilayer intra-LL
CE which consists of two excitations with in-phase and out-of-phase density
fluctuations and with the energies depending on the interlayer spacing and
the characteristics of individual 2DES.

The bilayer dielectric tensor $\varepsilon _{ij}(q,\omega )$ ($i,j=1,2$ are
the layer indices) is given in terms of the irreducible electron
polarization function $\Pi _{ij}(q,\omega )$, which is obtained from the
solution of a matrix Dyson equation for the dynamically screened Coulomb
interaction $\hat{V}(q,\omega )$. It has the following usual form
\begin{equation}
\hat{V}(q,\omega )=\hat{v}(q)\left[ 1-\hat{v}(q)\hat{\Pi}(q,\omega )\right]
^{-1}\equiv \hat{v}(q)\hat{\varepsilon}^{-1}(q,\omega )
\end{equation}%
where all quantities are $2\times 2$ matrices, $\hat{v}$ denotes the bare
Coulomb interaction tensor. For simplicity we assume that the layers are
infinitely thin and use $v_{11}=v_{22}=v=2\pi e^{2}/\kappa _{0}q$ and $%
v_{12}=v_{12}=\exp (-q\Lambda )v$, where $\kappa _{0}$ is the static
dielectric constant and $\Lambda $ the interlayer spacing. We neglect
tunneling between the layers and adopt an approximation where the interlayer
polarization functions are zero, $\Pi _{12}=\Pi _{21}=0.\,$Then, the
components of the inverse dielectric tensor are given as $\varepsilon
_{11,22}^{-1}(q,\omega )=\varepsilon _{2,1}\left( q,\omega \right)
/\varepsilon _{\text{bi}}\left( q,\omega \right) $ and $\varepsilon
_{12,21}^{-1}(q,\omega )=v_{12}\left( q,\omega \right) /\varepsilon _{\text{%
bi}}\left( q,\omega \right) $. Here, $\varepsilon _{1,2}\left( q,\omega
\right) =1-v_{11}\Pi _{11,22}(q,\omega )$ and $\Pi _{11,22}(q,\omega )$ are
the screening and polarization functions in each layer. The bilayer
screening function $\varepsilon _{\text{bi}}(q,\omega )$ is the determinant
of the dielectric tensor, $\varepsilon _{\text{bi}}(q,\omega )=\varepsilon
_{1}(q,\omega )\varepsilon _{2}(q,\omega )-v_{12}^{2}(q,\omega )\Pi
_{11}(q,\omega )\Pi _{22}(q,\omega )$.

The RPA approximates the exact irreducible $\Pi (q,\omega )$ by its simple
bubble diagram with respect to the e-e interaction, which, for finite
magnetic fields, gives
\begin{widetext}
\begin{eqnarray}
\Pi (q,\omega )=\frac{1}{\pi \ell _{B}^{2}}\sum_{l,l^{\prime }=0}^{\infty
}Q_{ll^{\prime }}\left( t\right) \int\limits_{-\infty }^{\infty }\frac{dE}{%
\pi }f_{T}(E-E_{F})\text{Im }G_{l}^{R}(E)\left( \frac{G_{l^{\prime
}}^{A}(E-\hbar \omega )}{\left[ 1-\delta \gamma _{ll^{\prime
}}^{AA}(t,E,E-\hbar \omega )\right] \left[ 1-\delta \gamma
_{ll^{\prime
}}^{RA}(t,E,E-\hbar \omega )\right] }\right.   \label{eq2} \\
+\left. \frac{G_{l^{\prime }}^{R}(E+\hbar \omega )}{\left[ 1-\delta \gamma
_{ll^{\prime }}^{AR}(t,E,E+\hbar \omega )\right] \left[ 1-\delta \gamma
_{ll^{\prime }}^{RR}(t,E,E+\hbar \omega )\right] }\right) ,\text{ \ \ }%
\delta \gamma _{ll^{\prime }}^{ab}(t,E,E\pm \hbar \omega )=\frac{\Gamma
_{0}^{2}}{4}Q_{ll^{\prime }}\left( t\right) G_{l}^{a}(E)G_{l^{\prime
}}^{b}(E\pm \hbar \omega ).  \nonumber
\end{eqnarray}%
\end{widetext}For brevity we suppress the layer indices here, $G_{l}^{R,A}(E)
$ are the electron retarded and advanced Green's functions, dressed by e-i
interaction. In the SCBA they are diagonal and given by $G_{l}^{R,A}(E)=2%
\left( E-E_{l}+\sqrt{\left( E-E_{l}\right) ^{2}-\Gamma _{0}^{2}}\right) ^{-1}
$ with the imaginary part of the square-root taken positive (negative) for
the advanced (retarded) function. In the short-range impurity model, the
half-width $\Gamma _{0}=\sqrt{\frac{2}{\pi }\hbar \omega _{B}\frac{\hbar }{%
\tau }}$ of the $E_{l}=(l+1/2)\hbar \omega _{B}$\ LL is independent of the
Landau index $l$ ($\tau $ is the transport relaxation time, determined from
mobility $\mu $). The bare vertex functions $Q_{ll^{\prime }}(t)$ are given
by the gauge invariant part of the in-plane form factor: $Q_{ll^{\prime
}}\left( t\right) =(-1)^{l+l^{\prime }}e^{-t}L_{l}^{l^{\prime }-l}\left(
t\right) L_{l^{\prime }}^{l-l^{\prime }}\left( t\right) $ where $t=(q\ell
_{B})^{2}/2$ and $L_{l}^{l^{\prime }}\left( t\right) $ is the associated
Laguerre polynomial. The denominators in the r.h.s. of Eq.~(\ref{eq2})
differ from one due to the vertex corrections $\delta \gamma _{ll^{\prime
}}^{ab}$ which are obtained consistently in the short-range impurity model
with the assumption that the LL are clearly resolved \cite{bons97}. The
chemical potential $E_{F}(n,B,T,\mu )$ in the Fermi distribution function $%
f_{T}$ is determined implicitly by the electron density via $n=-\left( 2\pi
^{2}\ell _{B}^{2}\right) ^{-1}\sum_{l=0}^{\infty }\int_{0}^{\infty
}dEf_{T}(E-E_{F})$Im$G_{l}(E)$. Here we use the electron Green's functions
which correspond to the Gaussian density of states, Im$G_{l}(E)=\sqrt{2\pi }%
/\Gamma _{0}\exp \left( -2\left( E-E_{l}\right) ^{2}/\Gamma _{0}^{2}\right) ,
$ without unphysical edges of the Landau band \cite{gerhardts}. We assume
spin degeneracy so the capacity of each LL is doubled.

\begin{figure}[t]
\includegraphics[angle=-90,width=8.5cm]{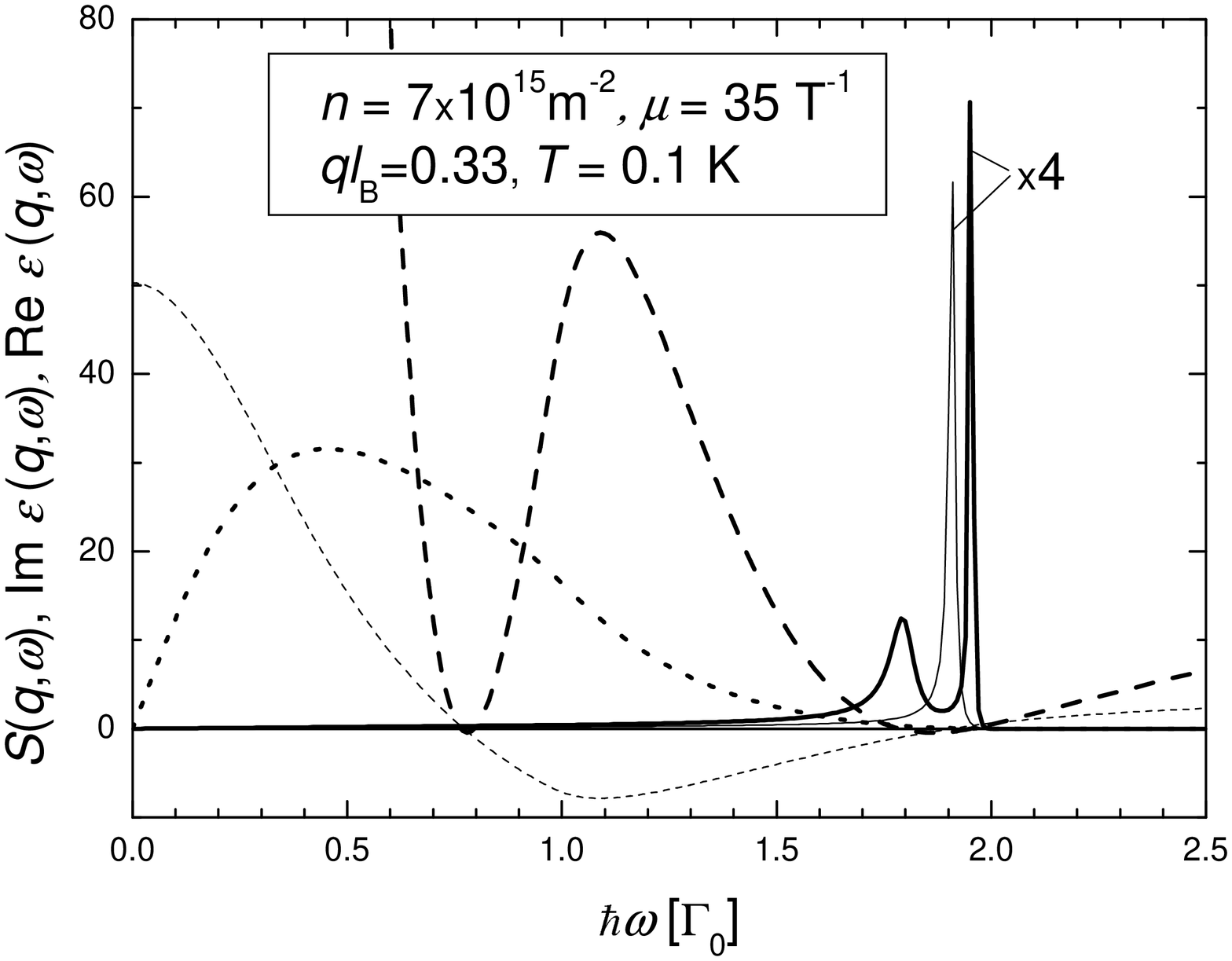}
\caption{The structure factor (solid line), the real (dashed line) and
imaginary (dotted line) parts of dielectric function of the bilayer 2DES
with the $\Lambda =30$ nm spacing and for $l_{0}=4$. The thin lines
correspond to the individual 2DES in which $\text{Im}~\protect\varepsilon (q,%
\protect\omega )$ differs from that of in the bilayer 2DES by a numerical
factor and is not shown here.}
\label{fg1}
\end{figure}

It is seen in Fig.~\ref{fg1} that $\text{Re}~\varepsilon (q,\omega )$ of the
individual 2DES becomes negative for intermediate values of $\omega $ of the
order of $\Gamma _{0}$. This requires two crossings of the $\text{Re}%
~\varepsilon (q,\omega )=0$ axis. The low-frequency zeros (LFZ) of $\text{Re}%
~\varepsilon (q,\omega )$ always lie within the PHC where $\text{Im }%
\varepsilon (q,\omega )$ is large. The high-frequency zeros (HFZ) are close
to or above the upper edge of the PHC where $\text{Im }\varepsilon (q,\omega
)$ is small or exactly zero within the SCBA. The pronounced peaks of the
dynamical structure factor $S(q,\omega )\propto -\text{Im}\left(
1/\varepsilon (q,\omega )\right) $ as a function of $\omega $ correspond to
HFZ, while no peaks correspond to LFZ. Therefore, HFZ represent intra-LL CE
of the system while LFZ cannot be interpreted as excitations. We believe,
however, that these particle-hole states play an important role,
particularly in realization of the frictional magnetodrag effect, and we
calculate LFZ and present them below in the spectrum together with the
intra-LL CE modes.

In the bilayer 2DES, there appears a pair of zeros instead of LFZ and HFZ of
the single layer 2DES and Re$~\varepsilon (q,\omega )$ becomes negative only
within small regions, restricted by the two zeros in each pair. Therefore,
the bilayer intra-LL CE spectrum consists of two branches which correspond
to the in-phase and out-of-phase intra-LL CE. Accordingly, one can see from
Fig.~\ref{fg1} that the bilayer structure factor, instead of the single peak
of $S(q,\omega )$ of the individual 2DES, shows a double-peak structure near
the upper edge of the PHC, while again it demonstrates monotonic behavior at
low energies.

\begin{figure}[t]
\includegraphics[angle=-90,width=8.5cm]{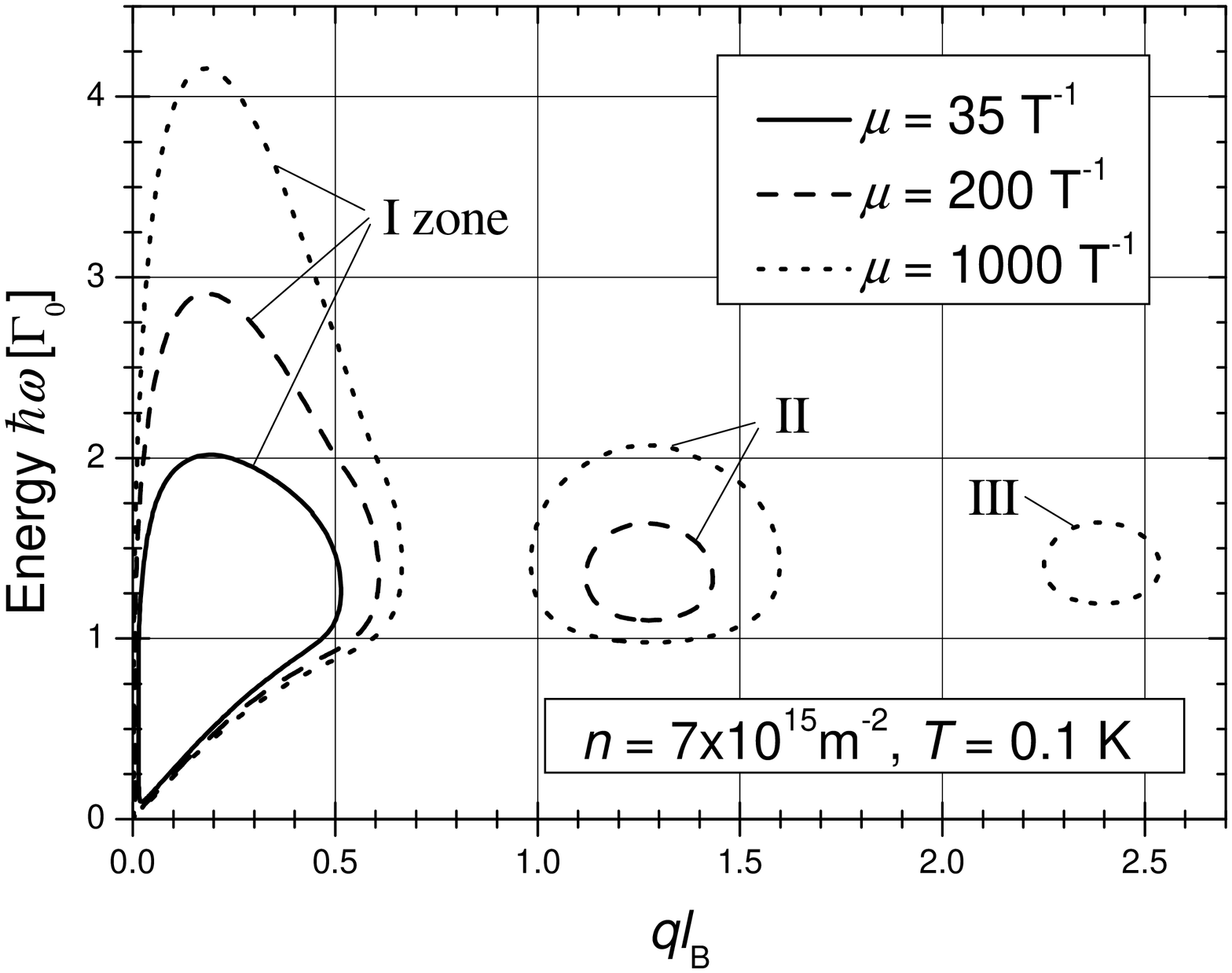}
\caption{The dispersion relations of intra-LL CE referred to the $l_{0}=4$
half-filled LL in the individual 2DES for the three values of mobility.}
\label{fg2}
\end{figure}

In Fig.~\ref{fg2} we plot the spectra of intra-LL CE of individual 2DES at
finite temperatures and for various values of electron mobility. Inclusion
of the consistent vertex corrections allows us to reproduce the square-root
like dispersion \cite{anton} of intra-LL CE at small momenta. However, no
exact $q\rightarrow 0$ limit exists in the spectrum. In the \textit{%
disordered} 2DES the density of states remains finite in the centrum of the
LL and the response function vanishes as $q^{2}$, even if $\omega $ is in
the immediate vicinity of $\omega _{B}$. In the short wavelength limit,
again the \textit{disordered }2DES does not respond since the response
function vanishes exponentially at wavelengths much shorter than $\ell _{B}$%
. When the lowest LL is partially filled, there exists only one allowed wave
vector zone, and our findings for $l_{0}=0$ are in agreement with Ref.~%
\onlinecite{anton}. For CE referred to the higher LL, one can see from Fig.~%
\ref{fg2} that the number of allowed wave vector zones can be multiple and
increases with $\mu $. Near half-filling of the $l_{0}=4$ LL and at $T=0.1$
K there exist only one, two, and three allowed zones for $\mu =35,200$, and $%
1000$ T$^{-1}$, respectively. Notice that for $l_{0}=4$ the first two of the
maximum four available wave vector gaps are located around $q_{r}\ell
_{B}=0.8$ and $1.87$. In the first zone close to $q=0$, for all three values
of $\mu $, the intra-LL CE emerge above the PHC in the middle of the zone
where the energy shows a maximum. In this region the spectrum describes CE
modes without damping and delta-function like peaks occur in $S(q,\omega )$.
Towards the edges of the allowed zone the energy decreases. Within the PHC $%
\hbar \omega <2\Gamma _{0}$ near its upper edge, the peaks of $S(q,\omega )$
become broadened, and though the intra-LL CE modes are damped but still are
well defined excitations. Exactly at the edges of the allowed zone, the
group velocity shows anomalous behavior and becomes infinite. At these
points CE merge with the particle-hole states, which are described by LFZ of
$\text{Re}~\varepsilon (q,\omega )$. Below this merging energy the damping
is so strong that the zeros of $\text{Re }\varepsilon (q,\omega )$ cannot be
interpreted as CE. The greatest maximum of energy and the largest width of
the allowed wave vector zone are achieved, when the outermost LL is near
half-filling, $\mu $ large, and $T$ low. In the higher zones the energy
maximum is smaller and the intra-LL CE emerge above the PHC only in the
second zone for the highest value of mobility.

\begin{figure}[tbp]
%[h,t,b,p]
\centering\includegraphics[angle=-90, width=8.5cm]{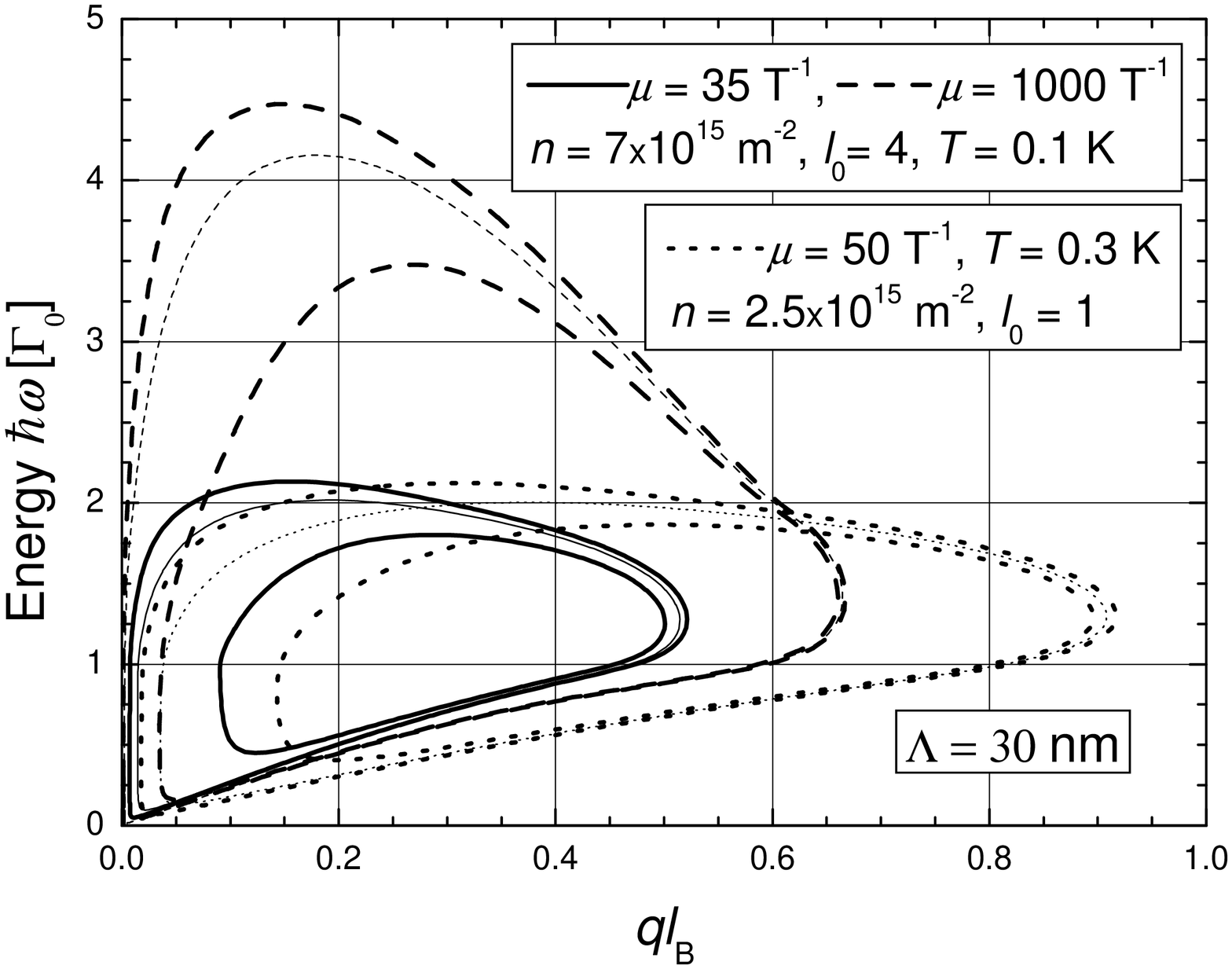}
% Here is how to import EPS art
\caption{The dispersion relations of intra-LL CE in the first allowed wave
vector zone for the bilayer 2DES. The thin lines represent the spectra in
the single-layer limit.}
\label{fg3}
\end{figure}

In Fig.~\ref{fg3} we plot the dispersion relations of intra-LL CE of the
bilayer disordered 2DES with the interlayer separation $\Lambda =30$ nm. We
restrict our consideration here to the symmetric bilayer 2DES with matched
electron densities and present calculations only for the first allowed wave
vector zone. In this case $\varepsilon _{1}=\varepsilon_{2}=\varepsilon $
and the bilayer screening function $\varepsilon _{\text{bi}}(q,\omega )$ is
represented as
\begin{equation}
\varepsilon _{\text{bi}}(q,\omega )=(1-\alpha ^{2})\left( \varepsilon
(q,\omega )-\varepsilon _{\text{in}}(q)\right) \left( \varepsilon (q,\omega
)-\varepsilon _{\text{out}}(q)\right)
\end{equation}%
where we have defined $\varepsilon _{\text{in,out}}(q)=1/(1\pm \alpha ^{-1})$
with $\alpha =\exp (-q\Lambda )$. Now the spectrum of intra-LL CE in the
bilayer 2DES is determined implicitly by the conditions $\text{Re }%
\varepsilon (q,\omega _{\text{in,out}}\left( q\right) )=\varepsilon _{\text{%
in,out}}(q)$.

As seen from Fig.~\ref{fg3}, the spectrum of bilayer intra-LL CE is split
into the in- and out-of-phase modes. The in-phase CE mode, irrespective of
the spacing $\Lambda $, is less affected by the interlayer interaction, lies
above and behaves similarly to the CE mode of individual 2DES. The
out-of-phase CE branch lies below the single layer mode and has a dipole
oscillator strength. At small wave vectors the dispersion of in-phase modes
is close to the square-root one while the out-of-phase modes demonstrate a
smoother dispersion and achieve the maximum energy at larger values of $q$.
The splitting energy, $\delta E$, decreases in $\Lambda $ which supports the
tendency that within the limit of $\Lambda \rightarrow \infty $, the
dispersion curves of in- and out-of-phase modes degenerate and coincide with
that of single layer intra-LL CE mode. $\delta E$ has a minimum near the
upper edge of the PHC and close to the right edge of the allowed wave vector
zone. Towards the right edge of the zone, $\delta E$ increases slightly,
while the increase is strong towards the left edge. The CE modes with $q$
close to the edges of the zone and with energies $\hbar \omega <2\Gamma _{0}$
are responsible for the enhancement of the frictional magnetodrag effect.
The out-of-phase CE mode has relatively strong damping and contributes
largely to the drag effect. However, in certain ranges of wave vectors close
to the edges of the allowed zone, the in-phase modes develop and exist
solely and will contribute to the drag effect in the absence of out-of-phase
modes.

The corresponding bilayer structure factor for the $l_{0}=4$ LL is depicted
in Fig.~\ref{fg1}, which we calculate in the plasmon-pole approximation. For
$q\ell _{B}=0.33$ the splitting energy is approximately equal to $0.17\Gamma
_{0}$. For $\mu =35$ T$^{-1}$ and $B\approx 3.22$ T we have $\Gamma
_{0}\approx 4.84$ K and this gives $\delta E$ more than $0.8$ K. This is a
quite measurable quantity, however, notice that at smaller wave vectors, $%
\delta E$ is appreciably larger (Fig.~\ref{fg3}). As seen from Fig.~\ref{fg1}%
, the out-of-phase peak of $S(q,\omega )$ is about six times lower than the
in-phase mode peak. However, both the in- and out-of-phase CE modes have
enough weight to be observed experimentally, and the signatures of these
bilayer intra-LL CE in the infrared absorption and inelastic light
scattering measurements should constitute asymmetric doublets.

We expect that the approximations we use here are adequate for the 2DES in
which disorder dominates over Coulomb correlations. This is safer in the
first allowed zone where $ql_{B}<1$ and the RPA is an acceptable
approximation even in the disorder free case \cite{kallin}. In the
disordered 2DES, the RPA is valid when no FQHE excitation gap is developed
\cite{anton}. This is the case if the gap in the disorder free case, $\Delta
_{0}$, is smaller than the typical disorder energy \cite{gap}. According to
the Laughlin's estimates \cite{laughlin} $\Delta _{0}\approx 0.056E_{c}$ for
the smallest fraction $\nu =1/3$, for which $\Delta _{0}$ is maximal. Here,
we consider moderate magnetic fields well below the FQHE threshold \cite%
{fqhe} and obtain the intra-LL CE referred to the higher LL. For $\mu =35$ T$%
^{-1}$, the ratio $E_{c}/\Gamma _{0}$ is less than $20$, and near
half-filling of the $l_{0}=4$ LL, the system is well away from the regime
where FQHE occurs. Thus, we expect that the regime considered here is
favorable for the applicability of this theory and for the observation of
these novel bilayer intra-LL CE with attractive many-body and disorder
effects.

\begin{acknowledgments}
We thank A. H. MacDonald, S. E. Ulloa, R. Gerhardts, and I. Kukushkin for
useful discussions. This work was supported by the Korea Science and
Engineering Foundation through Grant No. R05-2003-000-11432-0, by the
Ministry of Science and Education of Armenia under Grant No: 0103, and by a
DAAD technical grant.
\end{acknowledgments}


\begin{thebibliography}{99}
\bibitem{eisensuen} Y. W. Suen \textit{et al}., Phys. Rev. Lett. \textbf{68}%
, 1379 (1992); J. P. Eisenstein \textit{et al}., Phys. Rev. Lett. \textbf{68}%
, 1383 (1992).

\bibitem{isospin} K. Moon \textit{et al}., Phys. Rev. B \textbf{51}, 5138
(1995); K. Yang \textit{et al}., \textit{ibid}. \textbf{54}, 11 644 (1996);
L. Brey \textit{et al}., \textit{ibid}. \textbf{54}, 16 888 (1996); A. H.
MacDonald, R. Rajaraman, and T. Jungwirth, \textit{ibid}. \textbf{60}, 8817
(1999).

\bibitem{flensberg} K. Flensberg and B. Y.-K. Hu, Phys. Rev. Lett. \textbf{73%
}, 3572 (1994); Phys. Rev. B \textbf{52}, 14 796 (1995).

\bibitem{hill97} N. P. R. Hill \textit{et al}., Phys. Rev. Lett. \textbf{78}%
, 2204 (1997).

\bibitem{noh98} H. Noh \textit{et al}., Phys. Rev. B \textbf{58}, 12 621
(1998).

\bibitem{wu} M. W. Wu, H. L. Cui, and N. J. M. Horing, Modern Phys. Lett. B
\textbf{7}, 279 (1996).

\bibitem{khaetskii} A. V. Khaetskii and Yuli V. Nazarov, Phys. Rev. B
\textbf{59}, 7551 (1999).

\bibitem{manolescu} A. Manolescu and B. Tanatar, Physica E \textbf{13}, 80
(2002).

\bibitem{ssc} S. M. Badalyan and C. S. Kim, Solid State Commun. \textbf{127}%
, 521 (2003).

\bibitem{sarmachaplik} R. Z. Vitlina and A. V. Chaplik, Sov. Phys. JETP
\textbf{54}, 536 (1981); S. Das Sarma and A. Madhukar, Phys. Rev. B \textbf{%
23}, 805 (1981); G. E. Sontoro and G. F. Giuliani, Phys. Rev. B \textbf{37},
937 (1988).

\bibitem{kainth} D. S. Kainth \textit{et al}., Phys. Rev. B \textbf{59},
2095 (1999); A. Pinczuk, M. G. Lamont, and A. C. Gossard, Phys. Rev. Lett.
\textbf{56}, 2092 (1986); G. Fasol \textit{et al. ibid}. \textbf{56}, 2517
(1986).

\bibitem{aizin} G. R. A\u{\i}zin and G. Gumbs, Phys. Rev. B \textbf{52},
1890 (1995).

\bibitem{tovstonog} S. V. Tovstonog \textit{et al}., Phys. Rev. B \textbf{66}%
, 241308(R) (2002).

\bibitem{girvin} S. M. Girvin, A. H. MacDonald, and P. M. Platzman, Phys.
Rev. B \textbf{33}, 2481 (1986).

\bibitem{rennamd94} S. R. Renn and B. W. Roberts, Phys. Rev. B \textbf{48},
10926 (1993); A. H. MacDonald and S. C. Zhang, \textit{ibid}. \textbf{49},
17208 (1994).

\bibitem{kallin} C. Kallin and B. I. Halperin, Phys. Rev. B \textbf{30},
5655 (1984).

\bibitem{anton} D. Antoniou, A. H. MacDonald, and J. C. Swihart, Phys. Rev.
B \textbf{41}, 5440 (1990).

\bibitem{merkt} U. Merkt, Phys. Rev. Lett. \textbf{76}, 1134 (1996).

\bibitem{kohn} W. Kohn, Phys. Rev. \textbf{123}, 1242 (1961).

\bibitem{ando} T. Ando, A. Fowler, and F. Stern, Rev. Mod. Phys. \textbf{54}%
, 437 (1982); T. Ando and Y. Uemura, J. Phys. Soc. Jpn. \textbf{36}, 959
(1974).

\bibitem{bons97} M. C. B\o nsager \textit{et al}., Phys. Rev. B \textbf{56},
10314 (1997).

\bibitem{gerhardts} R. Gerhardts, Z. Phys. B \textbf{21}, 275 (1975);
\textbf{21} 285 (1975).

\bibitem{gap} A. H. MacDonald \textit{et al}., Phys. Rev. B \textbf{33},
4014 (1986).

\bibitem{laughlin} R. B. Laughlin, Surf. Sci. \textbf{142}, 163 (1985).

\bibitem{fqhe} G. S. Boebinger \textit{et al}., Phys. Rev. Lett. \textbf{55}%
, 1606 (1985).
\end{thebibliography}
\end{document}